\documentclass{ws-rv961x669}
\usepackage[square]{ws-rv-van}     % numbered citation/references (default)
\usepackage{ws-rv-thm}     % comment this line when `amsthm / theorem / ntheorem` package is used
\usepackage{subfigure}     % required only when side-by-side / subfigures are used
\makeindex
\usepackage{epsfig,graphicx,psfrag,here}
\usepackage{hyperref}
\newcommand{\bda}{\begin{\displaymath}\begin{array}{rl}}
\newcommand{\eda}{\end{array}\end{displaymath}}
\newcommand{\be}{\begin{equation}}
\newcommand{\ee}{\end{equation}}
\newcommand{\bdm}{\begin{displaymath}}
\newcommand{\edm}{\end{displaymath}}
\newcommand{\bea}{\begin{eqnarray}}
\newcommand{\eea}{\end{eqnarray}}
\newcommand{\no}{\nonumber \\}
\newcommand{\fs}{\; .}
\newcommand{\co}{\; ,}

\newcommand{\QCD}{\mbox{\tiny Q\hspace{-0.05em}CD}}

\newcommand{\MSbar}{$\mbox{MS}\hspace{-1.5em}\rule[0.8em]{1.5em}{0.05em}$ }

\newcommand{\al}{&\hspace{-0em}}

\newcommand{\ma}{\delta}
\newcommand{\pieta}{$\pi^0\!-\!\eta$ }

\def\eqref#1{(\ref{#1})}%

\begin{document}

\chapter{Testing $\chi$PT with the masses \\of the 
Nambu-Goldstone bosons\footnotemark}\footnotetext{In memory of Harald Fritzsch, contribution to the book edited by Gerhard Buchalla, Dieter L\"ust and Zhi-Zhong Xing.}

\author{H.~Leutwyler}
%\index[aindx]{Author, F.} % or \aindx{Author, F.}
%\index[aindx]{Author, S.} % or \aindx{Author, S.}

\address{Albert Einstein Center for Fundamental Physics\\Institute for Theoretical Physics, University of Bern\\
Sidlerstr.~5, CH-3012 Bern, Switzerland}

\begin{abstract}
The spontaneous breakdown of an approximate symmetry implies that the spectrum of the theory contains approximately massless particles. The hidden symmetry very strongly constrains their masses. A numerical evaluation of these constraints on the lattice would allow a more precise determination of the quark mass ratios $m_u:m_d:m_s$ and thereby reduce some of the uncertainties encountered in precision flavour physics.
\end{abstract}

%\markboth{Even Page Header}{Odd Page Header} % Customized running heads

\body

%\tableofcontents

\section{Introduction}
In 1972, Murray Gell-Mann gave a series of lectures at the Schladming Winter School.  At that time, the quarks were still taken as a theoretical construct to be treated like the veal used to prepare a pheasant in the royal french cuisine: the pheasant is baked between two slices of veal; while it deserves being served on the royal table, the veal stays in the kitchen for the less royal members of the court. Murray invited me to visit Caltech, which I did during three months in the spring break of 1973. There I met Harald Fritzsch and spent an extremely interesting period, collaborating with him and with Murray Gell-Mann on the possibility that the force between the quarks might be generated by an octet of gluons \cite{FGL}. 
The masses of the quarks is one of the puzzles which strongly attracted Harald's attention. During my visit at Caltech, we worked together on sum rules that involve the quark masses \cite{FL}. Later on, Harald wrote a paper  on his own \cite{Fritzsch 1977}, concerning the proposal that the Cabibbo angle might be related to quark mass ratios \cite{Gatto,Cabibbo,Oakes}. He intensively pursued this idea since then (for a recent account of his work in this direction see Ref.~\cite{Fritzsch Xing 2021}). 

It is not a simple matter to subject such relations to a stringent test, because the quark masses can be measured only indirectly, via their impact on measurable quantities. The lattice approach has made it possible to evaluate the Standard Model beyond perturbation theory. Slowly, but steadily the precision reached with these calculations is increasing. One of the problems encountered originates in the fact that the lightest hadron, the pion, is very light. The size of the box in which the system is enclosed must be large compared to the pion Compton wavelength for the presence of the box not to distort the properties of the system.  The fact that the e.m.~interaction is of long range represents an even more serious obstacle -- finite size effects persist even if the box is large. Currently, this appears to be the limiting factor in the determination of the quark mass ratios. 
Indeed, the current knowledge of the light quark masses is subject to substantial uncertainties. 

In the following, I try to draw attention to a theoretical question arising in this connection. The answer is within reach of presently available methods and could shed light on a poorly understood aspect of precision flavour physics: the sensitivity of the Standard Model predictions to the masses of the light quarks. 

As such, the quark masses do not represent physical quantities, but the renormalization procedure can be chosen such that their ratios do. It is customary to parametrize the two independent ratios of  the light quark masses with
\be\label{eq:RS} S\equiv\frac{m_s}{m_{ud}}\co\hspace{1cm}R\equiv\frac{m_s-m_{ud}}{m_d-m_u}\co\ee
where $m_{ud}\equiv\frac{1}{2}(m_u+m_d)$ stands for the mean mass of $u$ and $d$. 

The accuracy to which $S$ can be determined on the lattice is amazing. The results quoted in the most recent edition of the FLAG review  \cite{FLAG 2021} reads\footnote{Throughout the present article, the lattice numbers quoted stem from calculations with $N_f=2+1+1$.}:
\be\label{eq:Snum} S = 27.23(10)\,.\ee
The ratio $R$ is less well determined on the lattice because it concerns isospin breaking effects and is much more sensitive to the contributions from the e.m.~interaction than $S$.  Since this interaction is of long range, it is more difficult to properly account for on a lattice than QCD by itself.  While the above value of $S$ has an accuracy of about 4 per mille, the uncertainty attached to the result for $R$ in Ref.~\cite{FLAG 2021} is more than 10 times larger:
\be \label{eq:Rnum}R=35.9(1.7)\,.\ee

\section{QCD as part of the Standard Model}
The question I wish to draw attention to concerns QCD as such. I reserve the symbols $M_i$  for the physical masses and use $\hat M_i$  for what becomes of these if the electroweak interaction is turned off. It is well-known that the manner in which the electroweak interaction is turned off is not unique, but requires a convention, which can be specified as follows. I assume that  the Standard Model accurately describes nature unless the energies involved are too high.  At low energies, where the weak interaction is frozen, this model reduces to QCD + QED.  The neutrini can be described as free particles and the degrees of freedom of the intermediate bosons $W^\pm$, $Z$, the Higgs field and  the heavy quarks $b$, $t$ affect the low energy properties only indirectly. They are responsible for the fact that the neutron as well as the pions, kaons and myons decay. They do contribute to the masses of the particles and their magnetic moments, for instance, but these effects are much too small to be relevant at the accuracy reached in the determination of the quark masses. In the \MSbar scheme, the framework can thus be characterized by two running coupling constants and the running masses of four quarks and three charged leptons. 

As thoroughly discussed by Gasser, Rusetsky and Scimemi \cite{Gasser 2003}, the natural way to identify the QCD part of QCD + QED is to match the running coupling constant $\alpha_s(\mu)$ and the running quark masses $m_u(\mu)$, $m_d(\mu)$, $m_s(\mu)$, $m_c(\mu)$ of the two theories, at a suitable value of the renormalization scale $\mu$. Once the matching scale is chosen, the QCD part of the Standard Model is uniquely defined. Within QCD, the coupling constant and the quark masses do not run in the same manner as in the Standard Model and the masses $\hat M_i$ differ from the physical masses $M_i$ -- the difference represents the electroweak self energy.  Note also that QCD + QED involves contributions from virtual leptons. In the masses of the hadrons, these effects show up only at very high precision, but in principle, the matching of lattice calculations with the parameters of the Standard Model must account also for these contributions. 

The prescription to be used in the specification of the lattice version of QCD  is currently under critical examination within FLAG\footnote{Urs Wenger, private communication.}. I assume that, within errors, the numbers quoted in the FLAG review for $\hat M_{\pi^+}$, $\hat M_{K^+}$, $\hat M_{K^0}$, do represent the masses obtained within QCD from the values quoted for the quark masses. Since the low energy theorems discussed below hold irrespective of the values adopted for the coupling constant and the quark masses,  the matching with QCD + QED and with experiment does not play any role when comparing these predictions with lattice data. 
 
 \section{Expansion in powers of the quark masses}
 It so happens that three of the quarks are very light. If they were massless, QCD would have an exact chiral symmetry. The symmetry is partly hidden because the ground state is not symmetric with respect to chiral rotations. As pointed out by Nambu \cite{Nambu}, this implies that the spectrum of the theory contains massless particles: if the quarks were massless, the spectrum of QCD would contain an octet of massless Nambu-Goldstone bosons. 
 
 Chiral perturbation theory ($\chi$PT) treats the mass matrix of the light quarks, ${\cal M}=\mbox{diag}\{m_u,m_d,m_s\}$, as a perturbation and shows that, for the square of the masses of the charged pion and of the kaons, the expansion in powers of the light quark masses ("chiral expansion") starts with a linear term: 
\bea  \hat M_{\pi^+}^2\al =\al (m_u+m_d)B_0+O({\cal M}^2)\co\no
\hat M_{K^+}^2\al=\al (m_u+m_s)B_0+O({\cal M}^2)\co\label{eq:LO}\\
\hat  M_{K^0}^2\al=\al (m_d+m_s)B_0+O({\cal M}^2)\fs\nonumber\eea
On account of invariance under charge conjugation, these formulae also hold for $\pi^-$, $K^-$ and $\bar K^0$, but for $\pi^0$ and $\eta$, the leading terms are more complicated, because the difference $m_d-m_u$ breaks isospin and hence generates mixing between the two levels:
\bea \label{eq:LOpi0eta}\hat M_{\pi^0}^2\al=\al \left\{(m_u+m_d)-\mbox{$\frac{4}{3}$}(m_s-m_{ud})\sin^2\ma/\cos2\ma\right\} B_0+
O({\cal M}^2)\co\\
\hat M_{\eta}^2\hspace{0.15em}\al=\al \left\{\mbox{$\frac{2}{3}$}(m_{ud}+2m_s)+\mbox{$\frac{4}{3}$}(m_s-m_{ud})\sin^2\ma/\cos2\ma\right\} B_0+
O({\cal M}^2)\fs\nonumber\eea
 The mixing angle $\ma$ is determined by the quark mass ratio $R$ according to 
 \be \label{eq:tan}\mbox{tg}\, 2\,\ma=\frac{\sqrt{3}}{2R}\fs\ee
 Since $R$ happens to be large, $\delta$ is small. The repulsion of the two levels makes the neutral pion somewhat lighter than the charged one, but the effect is of second order in isospin breaking and hence very small. 
 
\section{Reparametrization invariance}
At leading order of the chiral expansion, the masses of the Nambu-Goldstone bosons do determine the ratios of the light quark masses, but the quark masses themselves cannot be determined within $\chi$PT: the reparametrization ${\cal M}'=\kappa {\cal M}$ of the quark mass matrix leaves the effective Lagrangian invariant, provided the low energy constant $B_0$ is transformed as well, with $B_0'=B_0/\kappa$. In the standard notation, the quark mass matrix enters the entire effective Lagrangian exclusively via the matrix $\chi=2B_0{\cal M}$, so that the higher order contributions are automatically  invariant under this transformation. 

At NLO, not even the quark mass ratios are determined by the meson masses. Kaplan and Manohar \cite{Kaplan Manohar} identified the algebraic origin of this property of $\chi$PT: it is a consequence of the fact that the effective Lagrangian only  exploits the symmetry properties of the quark mass matrix. The matrix ${\cal M}^{\dagger -1}\mbox{det}{\cal M}$ transforms in the same way under chiral transformations as ${\cal M}$ itself and the same thus holds for $ {\cal M}'={\cal M}+\lambda \,{\cal M}^{\dagger -1}\mbox{det}{\cal M}$.
The operation amounts to a reparametrization of the quark masses: $m_u'=m_u+\lambda \,m_d\, m_s$ and analogously for $m_d$ and $m_s$. Replacing ${\cal M}$ by ${\cal M}'$ in the effective Lagrangian leaves the leading order terms alone, but generates contributions of NLO that are quadratic in ${\cal M}$. The extra terms can be absorbed by changing the corresponding low energy constants (LECs) according to $L_6'=L_6-\bar \lambda$, $L_7'=L_7-\bar \lambda$, $L_8'=L_8+2\bar \lambda$, with $\bar \lambda=\lambda \,B_0/32 F_0^2$. 
The simultaneous change  ${\cal M}\rightarrow{\cal M}'$, $L_i\rightarrow L_i'$ does leave the effective Lagrangian invariant to NLO. Hence the chiral representation to first nonleading order of all quantities of physical interest obtained with this Lagrangian is  invariant under the above transformation. With a suitable transformation rule for the LECs occurring at higher orders, reparametrization invariance of the effective Lagrangian holds to all orders.

As a side remark, I mention that in the effective theory relevant if only two quark flavours are treated as light, the leading order Lagrangian only involves the sum of the two quark masses -- the difference only starts showing up at NLO. The Lagrangian is invariant under the reparametrization $m_u'+m_d'=\kappa(m_u+m_d)$, $m_d'-m_u' = \lambda (m_d-m_u)$, provided the LECs $B$, $\ell_7$ and $h_3$ are transformed with $B'=B/\kappa$, $\ell_7'=\kappa^2\ell_7/\lambda^2$, $h_3'=\kappa^2h_3/\lambda^2$.  In this framework, the quark mass ratio $m_u/m_d$ is not reparametrization invariant, either.

Since the quark mass ratios $S$ and $R$ are not reparametrization invariant, they do pick up NLO corrections that cannot be pinned down with $\chi$PT. Early estimates of the quark mass ratios \cite{Weinberg 1977,GL 1982} had to rely on LO formulae and were subject to considerable uncertainties that were often underestimated. An extreme example is the Dashen theorem \cite{Dashen}, which states that -- at leading order of the chiral expansion --  the e.m.~contributions to the square of the masses of the charged kaons and pions are the same, while the neutral Nambu-Goldstone bosons do not pick up such contributions at all. Langacker and Pagels \cite{Langacker:1973udm,Langacker:1978cf} pointed out even before $\chi$PT had been set up that the Dashen theorem neglects NLO contributions that contain juicy chiral logarithms. Several authors \cite{Maltman:1990mq,DHW 1993,Bijnens:1993ae,Urech,Baur:1995ig,Bijnens:1996kk,Ananthanarayan:2004qk} tried to estimate the low energy constants relevant at NLO, but the dust only settled when the work done on the lattice made it possible to solve QCD numerically. 

\section{\boldmath Meson mass ratio relevant for $S$}
When evaluating QCD on a lattice, the quark masses represent free parameters -- in principle, they can be taken arbitrarily small. In this limit, the ratios of the Nambu-Goldstone masses are given by ratios of quark masses. In particular, the ratio of the mean mass square  in the kaon multiplet, $\hat M_K^2=\frac{1}{2}(\hat M_{K^+}^2+\hat M_{K^0}^2)$, to the mass square of the charged pion is determined by the quark mass ratio $S$:  
\be \label{eq:DeltaS}\frac{2\hat M_K^2}{\hat M_{\pi^+}^2}=(S+1)(1+\Delta_S)\fs\ee
The factor $(1+\Delta_S)$  accounts for the corrections arising from higher orders of the expansion: $\Delta_S$ is of $O({\cal M})$.  

The uncertainties in the e.m.~self energies mainly concern the mass difference between the charged and neutral kaons. For the masses occurring in the above relation, the currently available lattice determinations imply 
\be \label{eq:Mnum} \hat M_{\pi^+}=134.8(3)\,\mbox{MeV}\,,\hspace{1em} \hat M_K=494.2(4)\,\mbox{MeV}\,.\ee 
Solving equation \eqref{eq:DeltaS} for $\Delta_S$ and using the value \eqref{eq:Snum} for $S$, I obtain
\be \label{eq:DeltaSnum}\Delta_S = -0.048(4)(3)\,,\ee
where the first and second error stem from the uncertainties in the meson and quark masses, respectively -- since the two are correlated, it is not legitimate to add them in quadrature. The result is displayed as a black dot with error bars in Fig.~1. It shows that the leading order prediction for the mass ratio $\hat M_K^2/M_{\pi^+}^2$ picks up remarkably small corrections from higher orders of the chiral expansion.

\begin{figure}[thb]\centering
\resizebox{0.8\textwidth}{!}{%
\includegraphics{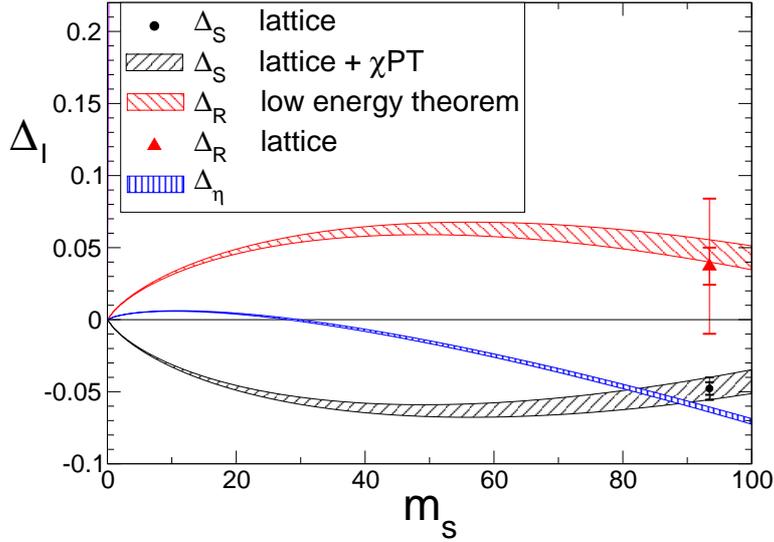}
}
\caption{Corrections to the leading order relations between the masses of the Nambu-Goldstone bosons and the quark masses. The ratios $m_u:m_d:m_s$ as well as  $\Lambda_{\QCD}$ and $m_c$ are kept fixed, $m_s$ is varied in the interval $0<m_s<100\,\mbox{MeV}$. The lattice values for $\Delta S$ and $\Delta R$ are based on the $N_f=2+1+1$ results quoted in the FLAG review \cite{FLAG 2021}. $\Delta_\eta$ represents the analogous correction to the Gell-Mann-Okubo formula.}    
\end{figure}

The chiral perturbation series for $\Delta S$ starts with
\bea\label{eq:DeltaSalg} \Delta S\al=\al (1-\delta_S)\left\{ -\mu_{\pi^0}+
\mu_\eta -\frac{6}{F_0^2}(\hat M_\eta^2-\hat M_{\pi^0}^2)(L_5^r-2L_8^r)\right\}+O({\cal M}^2)\,,\no
\delta_S\al=\al\frac{4(S+2) -2(S-1)\sec 2\hspace{0.05em}\ma}{3(S+1)}\,\sin^2\ma\co\hspace{1em}\mu_P\equiv\frac{\hat M_P^2}{32\pi^2 F_0^2}\ln \frac{\hat M_P^2}{\mu^2}\,.\eea
The term $\delta_S$ is of second order in isospin breaking and hence tiny: the quark mass ratios in equations \eqref{eq:Snum} and \eqref{eq:Rnum}  yield $\delta_S=1.11(10)\cdot 10^{-4}$. The first two terms in the curly bracket stem from loop graphs which are divergent and depend logarithmically on the masses of the Nambu-Goldstone bosons. The divergence is absorbed in a renormalization of the low energy constants $L_5^r$, $L_8^r$ and $\mu$ stands for the running scale used in the renormalization.  If isospin breaking is turned off $(m_u-m_d\rightarrow 0)$, the formula agrees with the representation for the masses of the Nambu-Goldstone bosons given in Ref.\cite{GL 1985}. 

The lower band in Fig.~1 illustrates the above formula by showing the dependence of $\Delta_S$ on the masses of the light quarks. Terms of  $O({\cal M}^2)$ are neglected and the combination $L_5^r-2L_8^r$ of LECs is fixed such that, at the physical value of  the quark masses\footnote{FLAG quotes $m_s=93.44(68)\,\mbox{MeV}$ in the $\overline{\mbox{MS}}$ scheme for $N_f=2+1+1$ \cite{FLAG 2021}.}, the band matches the linear sum of the errors in \eqref{eq:DeltaSnum}. Numerically, this leads to $L_5^r-2 L_8^r=-0.014(26)\cdot 10^{-3}$ at scale $\mu=M_\rho$, well within the range obtained from the individual numerical values of the LECs quoted in Ref.~\cite{FLAG 2021}. The strong curvature of the band illustrates the fact that the combination of LECs relevant here is very small -- the dependence on the quark masses is dominated by the chiral logarithms. 

The chiral expansion of the Nambu-Goldstone masses has been worked out explicitly to NNLO of the chiral expansion \cite{Amoros 2001,Anant 2018}. The package {\it Chiron} built by Hans Bijnens  \cite{Bijnens Chiron} includes everything needed to obtain the chiral representation for the massses of the Nambo-Goldstone bosons to two loops. The corresponding representation for the terms of $O({\cal M}^2)$ in formula \eqref{eq:DeltaSalg} involves further non-analytic terms as well as further LECs. The available numerical estimates of the latter are yet too crude to shed light on the size of the corresponding contributions to $\Delta S$, but the framework provides an excellent basis for the analysis of the lattice data. In the left half of the figure, these terms are negligible, but towards the right their importance grows. 

The work done on the lattice drastically reduced the uncertainty to which the quark mass ratio $S$ can be determined from phenomenology. The value \eqref{eq:DeltaSnum} shows that lattice calculations also yield a sharp determination for $\Delta S$ at the physical point. An accurate evaluation of the quark mass dependence of $\Delta S$ and of the term $\Delta R$ to be discussed below would allow to reduce the uncertainties in the LECs relevant for the Nambu-Goldstone masses. I expect this evaluation to confirm that, in Fig.~1, the neglected higher orders are indeed very small, throughout the range shown in that figure.

\section{Meson mass ratio relevant for $R$}

The ratio $R$ compares the difference $m_s-m_{ud}$, which is responsible for the breaking of the eightfold way, with the difference $m_d-m_u$, which breaks isospin symmetry. The relations \eqref{eq:LO} imply that, at leading order in the chiral expansion, the ratio of the corresponding differences between the squares of the Nambu-Goldstone masses is the same:
\be\label{eq:DeltaR} \frac{\hat M_K^2-\hat M_{\pi^+}^2}{\hat M_{K^0}^2-\hat M_{K^+}^2}=R\,(1+\Delta_R)\fs\ee
The correction $\Delta_R=O({\cal M})$ again accounts for the higher order contributions. 

The leading term in the chiral expansion of $\Delta_R$ not only involves the same low energy constants as $\Delta_S$, but also the same chiral logarithms: 
\be\label{eq:DeltaRalg} \Delta_R=-(1-\delta_R)\,\Delta_S+O({\cal M}^2)\co\hspace{2em}\delta_R=\frac{2S}{4R^2(S+1)+S-1} \fs\ee
In the isospin limit, the term $\delta_R$ vanishes. The isospin breaking effect is of second order also in this case and hence tiny: the quark mass ratios in equations \eqref{eq:Snum} and \eqref{eq:Rnum} yield $\delta_R=3.74(33)\cdot 10^{-4}$. Up to numerically irrelevant contributions, the low energy theorem \eqref{eq:DeltaRalg} thus simplifies to $\Delta_R=-\Delta_S+O({\cal M}^2)$. The upper band in Fig.~1 shows the prediction obtained for $\Delta_R$ if the NNLO corrections  are neglected. 

In order to test the low energy theorem \eqref{eq:DeltaRalg}, we need an estimate for the kaon mass difference in QCD, which occurs in the definition \eqref{eq:DeltaR}  of $\Delta_R$. 
The FLAG review \cite{FLAG 2021} discusses the matter in terms of the parameter $\varepsilon$ that measures the size of the higher order contributions to the Dashen theorem, but the results for the kaon mass difference can be worked out from the information given in the quoted references.

\begin{center}
\begin{tabular}{|c|c|c|c||c||c|}\hline \rule[1em]{0em}{0em}\hspace{-0.3em}
 reference &BMW \cite{BMW 2014}&RM123 \cite{RM123 2017}&MILC \cite{MILC 2018}&average&$\eta\rightarrow 3\pi$ \cite{CLLP}\\
\hline
\rule[-0.5em]{0em}{1.6em}\hspace{-0.7em} $\hat M_{K^+}^2\!-\!\hat M_{K^0}^2\hspace{-0.5em}$&6.149(122)&5.947(151)&6.075(125) &6.072(76)&6.24(38)\\
\hline\end{tabular}

\vspace{0.3em}
Table 1. Results for the kaon mass difference in QCD  (in units of $10^{-3}\,\mbox{GeV}^2$).
\end{center}

 Table 1 shows that the lattice results are consistent within errors. Their average determines the mass difference to an accuracy of 1.2\%.  The last entry of the table indicates that the phenomenological determination based on $\eta$ decay is much less accurate, but also consistent with the outcome of the lattice calculations. 
 
Evaluating the term $\Delta_R$ with the masses in \eqref{eq:Mnum}, the average for $\hat M_{K^+}^2-\hat M_{K^0}^2$ in Table 1 and the lattice result for $R$ in \eqref{eq:Rnum}, I obtain 
\be \label{eq:DeltaRnum}\Delta_R=0.037(13)(47)\,.\ee
The first error stems from the uncertainty in the meson masses and is totally dominated by the one in the mass difference  $\hat M_{K^+}^2-\hat M_{K^0}^2$. The second error represents the uncertainty due to the fact that the value of $R$ is known only to an accuracy of about 5\%. 

In Fig.~1, the central value of $\Delta_R$ is marked with a triangle. The small error bar represents the uncertainties in the meson masses, while the large one is the linear sum of the two sources of uncertainty.  This shows that, even if the uncertainty in $R$ is ignored, the result is consistent with the prediction. For a proper test of the low energy theorem, however, lattice calculations are required that simultaneously evaluate $\hat M_{K^0}^2-\hat M_{K^+}^2$ and $R$, so that the correlation between the two can be accounted for. With the currently available information, not even the sign of $\Delta R$ can be verified.

\section{\boldmath  Low energy theorem for $Q$}
\label{sec:Low energy theorem}
For the product of the two meson mass ratios in equations \eqref{eq:DeltaS} and \eqref{eq:DeltaR}, the chiral expansion starts with  \cite{GL 1985}
\be\label{eq:DeltaQ} \frac{\hat M_K^2 (\hat M_K^2-\hat M_{\pi^+}^2)}{M_{\pi^+}^2(\hat M_{K^0}^2-\hat M_{K^+}^2)}=Q^2(1+\Delta_Q)\,,\ee
where $Q^2\equiv\frac{1}{2} R(S+1)$ represents a ratio of quark mass squares,
\be \label{eq:Qm}Q^2\equiv\frac{m_s^2-m_{ud}^2}{m_d^2-m_u^2}\,,\ee
and $\Delta_Q$ accounts for the contributions of higher order. 

The first determination of $Q$ was based on an analysis of the decay  $\eta\rightarrow 3\pi$ in the framework of $\chi$PT \cite{GL 1985C}. This process is of particular interest because it violates the conservation of isospin and is therefore sensitive to the difference between $m_u$ and $m_d$. If the e.m.~interaction is ignored, the transition amplitude is proportional to $m_d-m_u$. As shown by Bell and Sutherland \cite{Sutherland,Bell Sutherland}, the e.m.~contributions are suppressed: in contrast to the mass differences $M_{\pi^+}^2-M_{\pi^0}^2$ and $M_{K^0}^2-M_{K^+}^2$ which do pick up substantial contributions from the e.m.~self energies already at leading order, the expansion of the e.m.~contribution to the transition amplitude only starts at $O(e^2 {\cal M})$. Accordingly, the quark mass ratio $Q$ can be expressed in terms of measured quantities, to next-to-leading order of the chiral expansion. More than 35 years ago, the numerical evaluation led to
 $1/Q^2=1.9(3)\cdot 10^{-3}$  \cite{GL 1985C}, which corresponds to $Q=23.2(1.8)$. 

In the meantime, the calculation was improved, accounting for higher order contributions in the expansion in powers of the momenta by means of dispersion theory \cite{Kambor Wiesendanger Wyler,Anisovich Leutwyler} as well as within $\chi$PT \cite{Bijnens:2007pr}. Also, the effects generated by the e.m.~interaction were studied in detail \cite{Baur Kambor Wyler,DKM}. For a thorough discussion, I refer to Ref.~\cite{CLLP}, where the outcome for $Q$ is given as
$ Q=22.1(7)$ -- this confirms the one loop result of $\chi$PT and is more accurate.

The determination of $Q$ from $\eta$ decay relies on the assumption that the correction $\Delta Q$ is negligibly small. The critical term in relation \eqref{eq:DeltaQ} is the mass difference $\hat M_{K^0}^2-\hat M_{K^+}^2$. Table 1 shows that the work done on the lattice has led to a significantly more accurate value for this quantity. Using the average listed in this table and the lattice results \eqref{eq:Mnum} for $\hat M_{\pi^+}$ and $\hat M_K$, the left hand side of \eqref{eq:DeltaQ} can be evaluated rather accurately. The higher orders of the chiral series produce the correction factor
\be\label{eq:QSR} (1+\Delta_Q)=(1+\Delta_S)(1+\Delta_R)\,.\ee
Equation \eqref{eq:DeltaRalg} implies that the two factors on the right hand side of this relation nearly compensate one another:
\be \label{eq:DeltaQalg} \Delta_Q=\delta_R\Delta_S+O({\cal M}^2)\fs\ee
Assuming that $\Delta_Q$ is indeed small compared to $\Delta_S$ or  $\Delta_R$, I obtain
\be \label{eq:QNLO} Q=22.4(2)\fs\ee
With the lattice result \eqref{eq:Snum} for $S$, the relation $Q^2=\frac{1}{2}R(S+1)$ then leads to
\be\label{eq:RNLO}R=35.5(5)\fs\ee
The quoted errors account for the uncertainties in all of the variables that enter the calculation as input, but do not include an estimate for the neglected higher order contributions. The uncertainty in the kaon mass difference dominates --  it is of the order of $1\%$. 

Concerning evaluations of $Q$ on the lattice, there were discrepancies, not only between different lattice calculations but also between some of these and phenomenology (see the 2019 edition of the FLAG review \cite{FLAG 2019}), but the dust appears to have settled:  the most recent update of this review \cite{FLAG 2021} quotes the value
\be\label{eq:QFLAG} Q=22.5(5)\,,\ee 
which is in perfect agreement with the value \eqref{eq:QNLO} obtained from the assumption that the correction $\Delta_Q$ is negligibly small. Note, however that the comparison does not provide a significant test of this assumption. Using the lattice results not only for the left hand side of \eqref{eq:DeltaQ} but also for $Q$, I obtain 
\be \label{eq:DeltaQFLAG} \Delta Q=-0.011(13)(43)\,,\ee
where the first error reflects the uncertainties in the meson masses, while the second stems from the one in the lattice result for $Q$.
The central value is indeed small, but since the uncertainty is large, it is not yet excluded that $\Delta Q$ is of the same size as $\Delta S$ -- contrary to what is expected from $\chi$PT.

\section{\boldmath Second order isospin breaking effects}
\label{sec:Second order}
The quark mass ratio $Q$ is not strictly reparametrization invariant. Also, in the chiral counting of powers, where the three light quark masses count as small quantities of the same order, the low energy theorem for $Q$ does not strictly hold to NLO. The quantity $\Delta_Q$, which represents the difference between the meson and quark mass ratios occurring in \eqref{eq:DeltaQ}, contains the term $\delta_R \Delta_S$, which is of $O({\cal M})$. It so happens that $m_d-m_u$ is small compared to $m_s-m_{ud}$, so that $\delta_R$ is tiny.  As discussed above, the isospin breaking effects encountered in the low energy theorem for $Q$ are too small to be of physical interest, but they do complicate matters. 

The change needed to arrive at a version of the low energy theorem that strictly holds to NLO is very modest: it suffices to replace
the quark mass ratio $Q^2$ by 
\be\label{eq:Qtilde} \tilde Q^2\equiv\frac{m_s^2-\tilde m_{ud}^2}{m_d^2-m_u^2}\,,\hspace{2em}\tilde m_{ud}^2\equiv\mbox{$\frac{1}{2}$}(m_u^2+m_d^2)\,.\ee
The differences between the squares of the light quark masses, $m_u^2-m_d^2$, $m_d^2-m_s^2$, $m_s^2-m_u^2$ are reparametrization invariant modulo contributions of $O({\cal M}^4)$. Since $\tilde Q^2$ can be expressed in terms of these differences, it is reparametrization invariant up to terms of $O({\cal M}^2)$, in contrast to $Q$. 

In order to find the corresponding change in the meson mass ratio that enters the low energy theorem, it suffices to solve the leading order mass formulae \eqref{eq:LO} for the quark masses. Inserting the result in the expression for $\tilde Q^2$, the denominator remains the same as before, while in the numerator, the term $\hat M_{K}^4$ is replaced by the product $\hat M_{K^0}^2 \hat M_{K^+}^2$.  Hence the correction $\Delta_{\tilde Q}$ defined by \be\label{eq:DeltaQtilde}  \frac{\hat M_{K^0}^2 \hat M_{K^+}^2-\hat M_{\pi^+}^2 \hat M_{K}^2}{\hat M_{\pi^+}^2
(\hat M_{K^0}^2-\hat M_{K^+}^2)}= \tilde Q^2(1+\Delta_{\tilde Q})\,\ee
vanishes at leading order of the chiral expansion. Actually, inserting the chiral expansion of the meson masses to NLO, one finds that not only the contributions from the LECs but also the chiral logarithms occurring at NLO cancel out, so that the low energy theorem takes the simple form
\be\label{eq:QtildeOMM} \Delta_{\tilde Q}=O({\cal M}^2)\,.\ee

The only difference between $Q$ and $\tilde Q$ is that $m_{ud}$ is replaced by $\tilde m_{ud}$. In view of 
$\tilde m_{ud}^2-m_{ud}^2=\frac{1}{4} (m_d-m_u)^2$, the difference is of 
second order in isospin breaking and hence tiny. At the quoted accuracy, the lattice result \eqref{eq:QFLAG} also holds for $\tilde Q$ and the values of $Q$ and $R$ obtained from the assumption that $\Delta \tilde Q$ is neglibly small cannot be distinguished from those given in equations \eqref{eq:QNLO} and \eqref{eq:RNLO}, obtained by neglecting $\Delta Q$.  

\section{Gell-Mann-Okubo formula}

The quantities $\Delta_S$, $\Delta_R$ and $\Delta_Q$ can be compared with the higher order corrections arising in the case of the Gell-Mann-Okubo formula, which predicts the mass of the $\eta$ in terms of  those of the pions and kaons. At leading order and in the isospin limit, the prediction reads $\hat M_\eta^2=\frac{1}{3}(4\hat M_{K}^2 -M_\pi^2)$ \cite{Gell-Mann 1961,Okubo 1962}. For $m_u\neq m_d$, the LO mass formulae \eqref{eq:LO} and \eqref{eq:LOpi0eta} imply that the relation takes the form $\hat M_\eta^2=\frac{1}{3}(2\hat M_{K^+}^2+2\hat M_{K^0}^2+2\hat M_{\pi^+}^2-3\hat M_{\pi^0}^2)$. The higher orders of the chiral expansion generate a correction which I denote by $\Delta_\eta$:
\be \label{eq:Deltaeta}\hat M_\eta^2= \frac{1}{3}(4\hat M_{K}^2 +2\hat M_{\pi^+}^2-3\hat M_{\pi^0}^2)(1+\Delta_\eta)\fs\ee
The numerical value of the correction is determined by the masses of the Nambu-Goldstone bosons. I discuss the estimates used for these, in turn. 

Since the $\eta$ is electrically neutral and spinless, its e.m.~self energy is expected to be positive but very small, comparable to the one of the $K^0$. The lattice results in equation \eqref{eq:Mnum} and Table 1 yield $M_{K^0}^{\mbox{\tiny QED}}=0.35(40)$ MeV.  In my opinion, the estimate $M_\eta^{\mbox{\tiny QED}}=0.4(4)$ MeV is on the conservative side. The corresponding range for the mass of the $\eta$ in QCD reads:
\be\label{eq:Metanum}\hat M_\eta=547.5(4)\,\mbox{MeV}\fs\ee

The lattice results in equation \eqref{eq:Mnum} determine the values of  $\hat M_K^2$ and $\hat M_{\pi^+}^2$ to good precision. Since the difference $\hat M_{\pi^+}^2-\hat M_{\pi^0}^2$ is of second order in isospin breaking, it is tiny. The change in the value of $\Delta_\eta$ obtained if  $M_{\pi^0}^2$ is replaced by $M_{\pi^+}^2$ is totally negligible compared to the uncertainty in the one from the kaons
and from the $\eta$, which are of comparable size. 

These estimates determine  the value of the correction to high accuracy: 
\be\label{eq:Deltaetanum}\Delta_\eta=-0.062(2)\fs\ee
The result is small, comparable with the one obtained for $\Delta S$, which also represents a quantity of $O({\cal M})$. The outcome is not sensitive to the estimate used for the mass of the $\eta$ in QCD: if $\hat M_\eta$ is instead identified with the physical mass of the particle, the central value of $\Delta_\eta$ is replaced by $-0.061$.

The one loop representation of $\Delta_\eta$,
 \bea\label{eq:Deltaetaalg}\Delta_\eta\al=\al \frac{2}{3\hat M_\eta^2}\left\{\rule[-1.3em]{0em}{2.6em}\hspace{-0.5em}- 3 \hat M_{\pi^0}^2\mu_{\pi^0}+2\hat M_{\pi^+}^2\mu_{\pi^+}+2\hat  M_{K^+}^2\mu_{K^+}+2\hat  M_{K^0}^2\mu_{K^0} - 3\hat  M_{\eta}^2\mu_{\eta} \right.\no
\al\al\rule{3em}{0em}\left. -\frac{3(\hat M_\eta^2-\hat M_{\pi^0}^2)^2}{F_0^2 }(L_5^r -12L_7-6L_8^r)\right\}
 +O({\cal M}^2)\,,\eea
is similar to the one for $\Delta_S$, but the combination of LECs occurring in $\Delta_\eta$ is reparametrization invariant, while the one in $\Delta_S$ isn't. Indeed, the definition \eqref{eq:DeltaS} of $\Delta_S$ contains the quark mass ratio $S$, which is not reparameterization invariant, while the definition \eqref{eq:Deltaeta} of $\Delta_\eta$ exclusively involves meson masses.  

The corrections $\Delta_S$ and $\Delta_\eta$ are both independent of the renormalization scale. The representation \eqref{eq:DeltaSalg} for $\Delta_S$ manifestly exhibits this property: the scale dependence of the logarithmic contributions cancels against the one of the LECs. In the above representation for $\Delta_\eta$, scale independence is not manifest, but it does hold if the LO formulae for the meson masses are inserted (this is a good check of the algebra).  As such, it does not matter whether the chiral logarithms are evaluated  with the LO expressions for the meson masses or with the estimates given for the full masses in QCD, for instance -- the difference in the result for $\Delta_\eta$ is of $O({\cal M}^2)$ . In order to preserve scale independence in numerical evaluations of the loop integrals, however, these must be expressed in terms of the parameters that occur in the chiral Lagrangian at leading order. The masses of the mesons running around the loops are given by the leading order formulae \eqref{eq:LO}, \eqref{eq:LOpi0eta} and depend on the quark mass ratios $S$, $R$, which are left open. For definiteness, I fix the scale with $B_0=2.52$ GeV; this choice  ensures that, at the physical value of $m_s$, the meson masses occurring in the loop integrals differ from the lattice results  for the full masses in QCD by less than $3\%$.

The narrow blue band in Fig.~1 displays the dependence of $\Delta_\eta$ on the light quark masses, again at fixed ratios $m_u:m_d:m_s$. The relevant combination of LECs is varied such that the value of $\Delta_\eta$ at the physical point matches the range \eqref{eq:Deltaetanum}. The plot shows that the lattice results determine the correction to the Gell-Mann-Okubo formula to remarkable precision. In size, the correction $\Delta_\eta$ is comparable to $\Delta_S$ or to the prediction for $\Delta_R$. The curvature of the bands stems from the chiral logarithms -- the contributions from the LECs are linear in the quark masses.
 
\section{\boldmath Low energy theorem for $\hat M_{\pi^+}\!-\!\hat M_{\pi^0}$}

At leading order, the chiral representation of the five mass squares  $\hat M_{\pi^0}^2$,  $\hat M_{\pi^+}^2$, $\hat M_{K^+}^2$, $\hat M_{K^0}^2$, $\hat M_{\eta}^2$ involves a single low energy constant, $B_0$. At NLO, three additional parameters appear: $L_4-2L_6$, $L_5-2L_8$, $3L_7+L_8$.  In the meson mass ratios, the first one of these as well as $B_0$ cancel out. Hence there are two parameter free constraints among the meson masses, valid to NLO.  The first one of these is the low energy theorem discussed above, which states that --  to NLO of the chiral expansion -- the meson mass ratio on the left hand side of equation \eqref{eq:DeltaQtilde} is given by the quark mass ratio $\tilde Q^2$. It may be viewed as a prediction for the isospin breaking mass difference  $\hat M_{K^0}-\hat M_{K^+}$ in terms of  the quark mass ratio $\tilde Q$, the mean kaon mass $\hat M_K$ and $\hat M_{\pi^+}$. The second prediction instead concerns the mass difference $\hat M_{\pi^+}-\hat M_{\pi^0}$, which represents an  isospin breaking effect as well, but while $\hat M_{K^0}-\hat M_{K^+}$ is proportional to $m_d-m_u$,  $\hat M_{\pi^+}-\hat M_{\pi^0}$ is of order $(m_d-m_u)^2$ and hence much smaller.

The chiral perturbation series for $\hat M_{\pi^+}-\hat M_{\pi^0}$ was worked out to NLO already in Ref.~\cite{GL 1985}, where a low energy theorem was established that relates the mass splittings in the two multiplets:
\be\label{eq:Deltapi} \hat M_{\pi^+}^2-\hat M_{\pi^0}^2=\frac{(\hat M_{K^0}^2-\hat M_{K^+}^2)^2}{3(\hat M_\eta^2-\hat M_{\pi^0}^2)}(1+\Delta_\pi)\fs\ee 
An explicit representation for $\Delta_\pi$ in terms of meson masses, valid to NLO of the chiral expansion, was given in the limit $m_u,m_d\rightarrow 0$, where the algebra simplifies considerably. In the remainder of the present section, this limitation is removed. 

Unfortunately, the fact that \pieta mixing occurs already at leading order of the chiral expansion leads to formulae at NLO that are too clumsy to be displayed here, but the numerical evaluation is straightforward and can be carried through without neglecting higher order isospin breaking effects. 

Equation \eqref{eq:Deltapi} defines $\Delta_\pi$ in terms of meson mass ratios. Inserting the leading order mass formulae  \eqref{eq:LO}, \eqref{eq:LOpi0eta}, one readily checks that $\Delta_\pi$ vanishes at LO of the chiral expansion. At NLO, the chiral representation of $\Delta_\pi$ involves a combination of low energy constants. Remarkably, the combination is the same as the one occurring in $\Delta_\eta$, i.e.~in the NLO correction to the Gell-Mann-Okubo formula. Eliminating the LECs in favour of the scale invariant quantity $\Delta_\eta$ with equation \eqref{eq:Deltaetaalg}, we thus arrive at a representation for $\Delta_\pi$ that holds to NLO of the chiral expansion and  exclusively contains the quark mass ratios $S$, $R$ and $\Delta_\eta$ (in particular, the prescription used for the evaluation of the chiral logarithms given in the preceding section only involves $S$ and $R$).

The available lattice results  restrict the quantities $S$, $\hat M_{\pi^+}$, $\hat M_K$, $\hat M_{K^0}^2-\hat M_{K^+}^2$, $\hat M_\eta$ to a narrow range. I evaluate the correction $\Delta_\pi$ on this basis, treat these quantities as known input and vary them in the range specified in equations \eqref{eq:Snum}, \eqref{eq:Mnum}, \eqref{eq:Metanum}  and Table 1. The low energy theorem \eqref{eq:QtildeOMM} implies that --  to NLO of the chiral expansion -- the quark mass ratio $\tilde Q$ is given by a ratio of meson masses that belong to the list of input variables.  Since the ratio $R$ can be expressed in terms of $\tilde Q$ and $S$, this quantity is also determined. The definition \eqref{eq:Deltaeta} of $\Delta_\eta$, however, involves the mass of the neutral pion, which is not contained in that list (in the evaluation of $\Delta_\eta$ discussed in the preceding section, higher order isospin breaking effects were neglected -- this is why the problem did not arise there). For the numerical analysis, however, it makes little difference whether $\hat M_{\pi^0}$ only occurs on the left hand side of formula \eqref{eq:Deltapi} or also on the right hand side -- with the NLO representation for $\Delta_\pi$, this formula determines the value of $\hat M_{\pi^0}$ in terms of known input in either case. Numerically, I obtain
\be\label{eq:Mpi0num}\hat M_{\pi^0}=134.58(30)\;\mbox{MeV}\co\hspace{2em}\Delta_\pi = 0.339(17)\fs\ee 
The errors account for the uncertainties in the input variables, but do not include an estimate for the neglected higher order contributions.
 The uncertainty in $\hat M_{\pi^0}$ is totally dominated by the one in $\hat M_{\pi^+}$ -- the uncertainty in the mass difference is much smaller:
 \be\label{eq:MDpinum}\hat M_{\pi^+}-\hat M_{\pi^0}=0.217(6) \;\mbox{MeV}\fs\ee
 
The central value $\hat M_{\pi^+}-\hat M_{\pi^0}=0.17(3)$ MeV obtained in Ref.~\cite{GL 1985} is lower and the uncertainty attached to it is much larger. The difference arises because the Dashen theorem was used to estimate the mass difference $\hat M_{K^0}^2-\hat M_{K^+}^2$ -- the work done on the lattice has shown that this theorem receives large corrections from higher orders of the chiral expansion. These are now known quite accurately.  
 
I emphasize that the above calculation ignores higher order contributions -- the quoted error exclusively accounts for the noise in the input. The outcome for $\Delta_\pi$ shows that, in this case, the low energy theorem  receives a large correction at NLO. In view of this, the error given in \eqref{eq:MDpinum} underestimates the present uncertainty in the mass difference: the neglected higher order contributions may well be larger than those arising from the errors attached to the lattice results used in the evaluation of the first two terms of the chiral perturbation series. The low energy theorem for $\tilde Q$ is on an entirely different footing: it does not receive  NLO corrections at all. A lattice calculation within QCD is needed to reliably determine $\hat M_{\pi^+}-\hat M_{\pi^0}$ to comparable accuracy. 

\section{Discussion, summary and conclusion}
1.~At leading order of the chiral expansion, the $\chi$PT formulae for the masses of the Nambu-Goldstone bosons in QCD provide a crude estimate for the ratios of the light quark masses. At NLO, these formulae involve low energy constants that cannot be determined in this way. In particular, the LO relation between the meson mass ratio $\hat M_K^2/M_{\pi^+}^2$ and the quark mass ratio $S=m_s/m_{ud}$ picks up a correction from nonleading orders of the chiral expansion,  measured by the term $\Delta_S$ defined in equation \eqref{eq:DeltaS}. The available lattice results show that this correction is remarkably small.

2.~The quark mass ratio $R=(m_s-m_{ud})/(m_d-m_u)$ is known much less well because it concerns the isospin breaking mass difference $m_d-m_u$,  which is small and not easy to disentangle from the isospin breaking effects generated by the e.m.~interaction.  The low energy theorem for $Q^2\equiv\frac{1}{2}R(S+1)$, however,  correlates $R$ with $S$: the term $\Delta_Q$ defined in \eqref{eq:DeltaQ} is strongly suppressed. At NLO of the chiral expansion, it represents a second order isospin breaking effect that is negligibly small compared to $\Delta_S$. 

3.~The quark mass ratio $Q$ is not reparametrization invariant, but a modest variation suffices to repair this shortcoming. The quantity $\tilde Q$ defined in \eqref{eq:Qtilde} differs from $Q$ only through numerically irrelevant contributions of second order in isospin breaking. More importantly, expressed in terms of $\tilde Q$, the low energy theorem discussed in the preceding paragraph takes a very simple form: the chiral expansion of the meson mass ratio on the left hand side of \eqref{eq:Qtilde} agrees with $\tilde Q^2$, not only at LO, but also at NLO of the chiral expansion. This implies that, if the quark masses are taken sufficiently small, the term $\Delta_{\tilde Q}=O({\cal M}^2)$ is negligible compared to $\Delta_S=O({\cal M})$. The assumption that the physical quark masses are sufficiently small in this sense leads to a rather sharp prediction for the value of $Q$ and -- together with the lattice result for $S$ -- also for $R$:
\be Q=22.4(2)  \,,\hspace{2em}R=35.5(5)\,.\ee
A more precise determination of $\hat M_{\pi^+}$, $\hat M_{K^+}$ and $\hat M_{K^0}$  is needed to determine the size of $\Delta_{\tilde Q}$. Note that the issue concerns the relation between the meson masses and those of the quarks in QCD. The self energies generated by QED are not important here. What is needed to submit the low energy theorem $\Delta_{\tilde Q}=O({\cal M}^2)$  to a crucial test is a precise determination of the meson masses that belong to a given input for the quark masses within QCD.  
 
\begin{figure}[thb]\centering
\resizebox{0.76\textwidth}{!}{%
\includegraphics{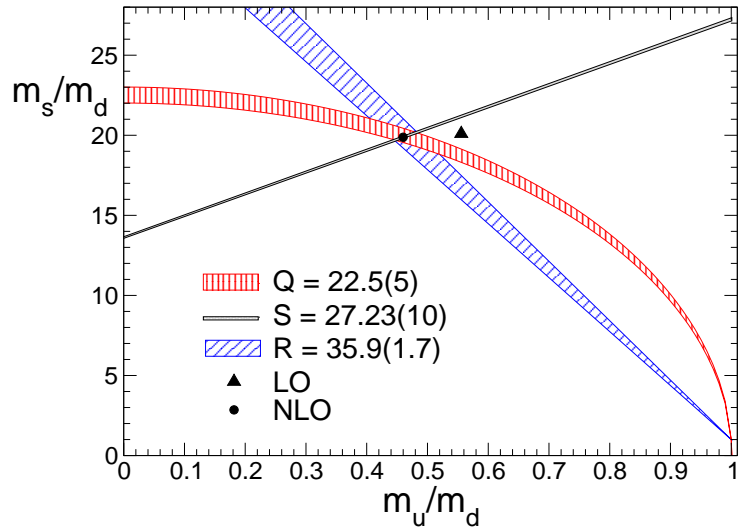}
}
\caption{Quark mass ratios. The numerical values of $S$, $R$ and $Q$ are taken from the entries for $N_f=2+1+1$ flavours in the FLAG review \cite{FLAG 2021}. The triangle shows the values obtained from Weinberg's leading order formulae  \cite{Weinberg 1977} for the mass ratios of the light quarks, which account for the e.m.~self energies with Dashen's theorem.}\end{figure}

4.~In the plane of the quark mass ratios $x=m_u/m_d$, $y=m_s/m_d$, a given value of $\tilde Q$ corresponds to an ellipse centered at $x=y=0$:
\be \frac{x^2}{a^2}+\frac{y^2}{b^2}=1\,,\hspace{1em} a^2=\frac{2\,\tilde Q^2+1}{2\,\tilde Q^2-1}\,,\hspace{1em}b^2=\tilde Q^2+\mbox{$\frac{1}{2}$}\,.\ee
While $a$ is very close to $1$, $b$ is large. The red band in Fig.~2 shows the region allowed by the lattice results for $Q$, which agree very well with those obtained from $\eta$ decay and are even somewhat more accurate. Visibly, the constraint imposed by the lattice results for $S$ is much stronger. It is indicated by a narrow wedge that intersects the $x$-axis at $x=-1$, close to the point $x=-a$, where the ellipse crosses this axis. 
The black dot marks the intersection of the lattice results for $S$ with the narrow band for $Q$ obtained by evaluating the chiral expansion of the corresponding ratio of meson masses to NLO and neglecting corrections of higher order.
Lines of constant $R$ instead pass through the point $x=y=1$, where the quark masses are the same and the eightfold way is an exact symmetry. $R$ is more sensitive to the effects generated by the e.m.~interaction than $S$. As these are of long range and hence more difficult to account for on the lattice, the corresponding wedge is considerably broader.

5.~It is not known why the strong interaction is CP-invariant to an extremely high degree of accuracy. If the $u$-quark were massless,  this puzzle would be solved. The fact that the one loop formulae of $\chi$PT only constrain the quark mass ratios to an ellipse, while the position on the ellipse depends on LECs that cannot be determined within $\chi$PT, led to the suggestion that the NLO corrections could be so large that they shift Weinberg's LO values  \cite{Weinberg 1977}, which in the figure are marked with a triangle, to the point on the ellipse that belongs to $m_u/m_d=0$. The ratios $S$ and $Q$ would then be related by $S^2=4\hspace{0.08em}Q^2+1$. With the lattice result for $Q$, this implies $S=45.0(1.0)$, more than 17 standard deviations away from the lattice result for $S$. This corroborates the conclusion reached long ago \cite{munotzero}: $m_u=0$ is an interesting way not to understand this world -- it is not the only one. 

6.~The triangle is rather close to the intersection of the three bands which represents the present knowledge of the quark mass ratios. In fact, if the Dashen theorem that underlies Weinberg's formulae is replaced by the present knowledge of the e.m~self energies, the triangle moves even closer to the physical point: the numerical value \eqref{eq:DeltaSnum} for $\Delta_S$ shows that the lattice result for $S$ deviates from the LO  formula $S+1=2\hat M_K^2/\hat M_{\pi^+}^2$ only by about 5\%.

7.~The Gell-Mann-Okubo formula also receives a correction at NLO. The explicit expression in \eqref{eq:Deltaetaalg} shows that the relevant combination of LECs is reparametrization invariant: in contrast to the definition \eqref{eq:DeltaS} of $\Delta_S$, the analogous formula \eqref{eq:Deltaeta} for $\Delta_\eta$ exclusively involves the meson masses. Numerically, $\Delta_S$ and $\Delta_\eta$ are of comparable size. 

8.~The states $\pi^0$ and $\eta$ undergo mixing. The repulsion of the two levels makes the $\pi^0$ lighter than the $\pi^+$. The shift is of second order in $m_d-m_u$ and hence small compared to the splitting generated by the e.m.~interaction. The low energy theorem \eqref{eq:Deltapi} relates the difference between the masses of the charged and neutral pions in QCD to the mass splitting in the kaon multiplet. The NLO correction $\Delta_\pi$ involves the same reparametrization invariant combination of low energy constants as the Gell-Mann-Okubo formula, but since it is large, the uncertainties arising from the neglect of the higher order contributions are considerable. 

\section*{Acknowledgments}
I thank Balasubramanian Ananthanarayan, Hans Bijnens, Gilberto Colangelo, J\"urg Gasser, Akaki Rusetski and Urs Wenger for useful comments on the manuscript.

\end{document}